\documentclass[hyper]{JHEP3} 

\usepackage{graphicx}
\usepackage{epsfig}
\usepackage{cite}
\usepackage{amsmath}

\newcommand{\ml}{{m_{\nu_1}}}
\newcommand{\mll}{{m^2_{\nu_1}}}
\newcommand{\mr}{{m_{n_1}}}
\newcommand{\mrr}{{m^2_{n_1}}}
\newcommand{\mh}{{m_h}}
\newcommand{\mz}{{m_Z}}

\newcommand{\met}{\not \!\! E_T}

\def\ie{{\it i.e.}}
\def\eg{{\it e.g.}}

\def\lsim{\mathrel{\mathpalette\atversim<}}
\def\gsim{\mathrel{\mathpalette\atversim>}}

\def\lsim{\mathrel{\rlap{\lower4pt\hbox{\hskip1pt$\sim$}}
    \raise1pt\hbox{$<$}}}         
\def\gsim{\mathrel{\rlap{\lower4pt\hbox{\hskip1pt$\sim$}}
    \raise1pt\hbox{$>$}}}         
\newcommand{\eeq}{\end{eqnarray}} 

\title{A vector-like fourth generation with a discrete symmetry from Split-UED}

\author{Kyoungchul Kong\\
        Theoretical Physics Department, SLAC, 
        Menlo Park, CA 94025, USA \\
        E-mail: \email{kckong@slac.stanford.edu}
        }

\author{Seong Chan Park\\
	Institute for the Physics and Mathematics of the Universe, \\
        The University of Tokyo, Kashiwa, Chiba 277-8568, JAPAN\\
	E-mail: \email{seongchan.park@ipmu.jp}
        }

\author{Thomas G. Rizzo\\
        Theoretical Physics Department, SLAC, 
        Menlo Park, CA 94025, USA \\
        E-mail: \email{rizzo@slac.stanford.edu}
        }

\preprint{
          IPMU10-0054 \\
          SLAC-PUB-14049 \\
          \today     }	

\abstract{
Split-UED allows for the possibility that the lowest lying KK 
excitations of the Standard Model fermions can be much lighter than the 
corresponding gauge or Higgs KK states. This can happen provided the fermion 
bulk masses are chosen to be large, in units of the inverse compactification 
radius, $1/R$, and negative. In this setup, all of the other KK states would 
be effectively decoupled from low energy physics. Such a scenario would 
then lead to an apparent vector-like fourth generation with an associated 
discrete symmetry that allows us to accommodate a dark matter candidate. In 
this paper the rather unique phenomenology presented by this picture will be 
examined.}


\begin{document}

\section{Introduction}
\label{sec:intro}

Although expectations for new physics beyond the Standard Model (SM) are 
ubiquitous due to the gauge hierarchy, flavor and dark matter problems, no 
one knows what form this new physics may take.  However, now a new window 
has opened on the Terascale: the long anticipated LHC has begun collecting 
data at $\sqrt s=7$ TeV. 

For almost a dozen years, models with extra spatial dimensions have been a 
popular way to address at least some of these outstanding problems \cite{ADD1, RS}. 
Among these, the Universal Extra Dimensions (UED) scenario \cite{Appelquist:2000nn} is 
particularly interesting as it leads to a dark matter (DM) candidate and can be 
embedded into the Large Extra Dimensions model to address the hierarchy problem\footnote{Assuming $d=1+N$ extra spatial dimensions  among which only one dimension ($\sim R$) is probed by matter fields (gauge bosons and fermions) but other dimensions ($\sim r$) only by graviton, we get the fundamental gravity scale in $D=4+d$ dimensions, $M_D$, much lower than what is expected with one ${\rm TeV}^{-1}$ sized extra dimension. The relation $M_4^2 \simeq M_5^3 R = (M_D^{3+N}r^N)R$ leads to $M_D\gsim 1\,{\rm TeV}\sim R^{-1}$ provided that $r^{-1}\sim 10^{-30/N}\,{\rm TeV}$. }. In minimal UED (MUED) there is assumed to be  one extra (orbifolded) dimension with a radius $R \sim 1\,{\rm TeV}^{-1}$ \cite{Antoniadis, Cheng:2002ab}, the only other free parameter, $\Lambda$, being the cut-off scale needed to define the effective 5D theory. These two parameters not only determine the interactions in MUED at tree-level but also at 1-loop \cite{Cheng:2002iz} assuming the vanishing boundary condition at the cutoff scale. In particular, the masses and couplings of all of the Kaluza-Klein(KK) excitations of the fermions and gauge bosons depend only upon values of these two parameters. To leading order these KK masses are simply given by $n_{KK} R^{-1}$ where $n_{KK}$ labels the KK excitation level with a corresponding KK-parity $(-1)^{n_{KK}}$. 

A variant of MUED, Split-UED(SUED) \cite{sued1, sued2, sued3, sued4}, 
allows for the existence of bulk 
mass terms for the various SM fermions, which are consistent with both gauge 
invariance as well as 5D Lorentz invariance. In order to satisfy the 
$S^1/Z_2$ orbifold symmetry of the action and to maintain KK-parity these 
mass terms, \ie, the coefficients of the various $\bar \Psi \Psi$ fermion 
bilinears, must be odd functions of the 5D coordinate, 
$-L\leq y \leq L,~L=\pi R/2$. The simplest choice is to assume that these 
coefficients are of the form $\mu_i \theta(y)$ with the $\mu_i$ being 
generally different for each chiral SM fermions with values which are 
naturally of order $\sim R^{-1}$ and where $\theta(y)=1(-1)$ for $y>(<)0$. In 
such a scenario the KK fermion masses will now depend not only on the value 
of $R^{-1}$ but also on the choices made for the parameters $\mu_i$. 

In this SUED model something unusual is found to happen to the mass of the 
lowest fermion KK excitation when $\mu$ becomes large and negative, \eg, when 
$\mu L \leq -1$. 
Whereas one might naively expect such masses to be of order 
$\sim R^{-1}$, one finds that the masses become exponentially suppressed  
($m_1 \simeq 2|\mu|\, e^{-|\mu|L})$. This means that it is possible to realize an unusual 
scenario: we can imagine that the size of extra dimension is tiny $R^{-1} \gg  1$ TeV so that all of the KK gauge boson, KK Higgs boson and $n\geq 2$ fermion KK states are very massive, beyond the range of the LHC and decoupled from low energy physics. At the TeV scale the only observable new states are then the lowest KK fermions. With 
respect to the usual SM, these states are {\it not} chiral but appear as 
a vector-like fourth generation which still carries negative KK-parity 
allowing us to identify the lightest of them as a dark matter candidate. It 
is the phenomenology of this unusual scenario that we will analyze in the 
discussion below.

\section{Decoupling KK modes with $n_{KK} \geq 2$}
\label{sec:light}
In this section we will discuss the KK spectrum of SUED concentrating on the light KK mode in particular. It is well known that the light Kaluza-Klein states appear due to a large extra dimension as the mass gap is roughly given by $\sim 1/R$ where $R$ is the `size' of the compact dimension \footnote{However, there are well-known counter examples with two or more extra dimensions such as compact hyperbolic spaces and the 2D torus with degenerate complex structure.}. 
An interesting observation is that a light Kaluza-Klein state ($\sim {\rm TeV}$) can appear even when the size of extra dimension is tiny $R \ll 1\,{\rm TeV^{-1}}$, if an odd bulk mass, $m_5(-y)=-m_5(y)$, is allowed. 
 Indeed, an odd mass term has been introduced  recently \cite{sued1} to maintain Kaluza-Klein parity 
which guarantees the stability of Kaluza-Klein dark matter \cite{Servant:2002aq, Cheng:2002ej}.

Let us consider a fermion ($\Psi$) with an odd mass term ($m_5$). The action is given as:
\begin{equation} 
        \label{eq:BulkAction}
 S = \int d^4 x \int_{-L}^{+L} dy 
\left[\frac{i}{2}
\bar{\Psi}\, \Gamma^M \overleftrightarrow{\partial}_M \Psi
 - m_5 \bar{\Psi} \Psi  \right],
\end{equation} 
where $a \overleftrightarrow{\partial_M} b = a \partial_M b - (\partial_M a) b$, $\Gamma^M=(\gamma^\mu, i\gamma_5)$.
This action is invariant under the inversion around the center of the orbifold ($y=0$), 
provided $m_5(y) \to m_5(-y)= -m_5(y)$, dubbed {\it the Kaluza-Klein parity} transformation: $y \to -y$ under which $\Psi(x,y)\to \Psi(x,-y)=\pm \gamma_5 \Psi(x,y)$. To parametrize the intrinsically odd mass term, we introduce a kink-type mass as
$m_5 = \mu ~\theta(y)$, where $\theta(y)=1(-1)$ for $y>(<)0$.

The fermion field is formally decomposed into two parts as $\Psi(x,y) = \tfrac{1-\gamma_5}{2} \Psi(x,y) + \tfrac{1+\gamma_5}{2} \Psi(x,y)=\Psi^L+\Psi^R$. 
In general, when a fermion belongs to a complex representation of the symmetry group, the KK modes can only acquire Dirac masses and the KK decomposition is of the form
\begin{eqnarray} 
        \label{eq:DiracKK}
\Psi^{L/R}=\sum_n \psi^{L/R}_n (x) f^{L/R}_n (y), 
\end{eqnarray} 
where $n$ labels the KK-level and $\psi^{L/R}_n$ are 4D  spinors which satisfy coupled Dirac equations:
$i \gamma^\mu \partial_\mu \psi_n^{L/R}=m_n \psi_n^{R/L}$.
The Kaluza-Klein spectrum is obtained by solving the Harmonic equation of the basis function:%
\begin{eqnarray} 
0=(\partial_5^2 +\Delta_n^2)f^{R/L}_n,
 \label{eq:2nd}
\end{eqnarray} 
where $\Delta_n^2 (\equiv m_n^2 -m_5^2)$ is determined by the given set of boundary conditions.  In general, $\Delta_n^2$ can have either positive or negative sign. On the branch with the positive sign ($\Delta_n^2>0)$, dubbed the `heavy branch' as the KK mass is heavier than the bulk mass, an infinite tower of KK states exists. This case has been already discussed in detail in the context of cosmology \cite{sued1,sued2,sued3} and collider physics \cite{sued2, sued4}. On the other hand, on the branch with the negative sign ($\Delta_n^2=-\kappa_n^2<0$), dubbed the `light branch', in which we are interested in this paper, a unique KK state ($n_{KK}=1$) exists only for a certain range of $\mu L$: $\mu L \leq -1 ( \geq 1)$ when the Dirichlet boundary condition is imposed to the left (DL) (right (DR)) component, respectively. On this branch, the first KK mass is lighter than the bulk mass as  $m_n^2 =\mu^2 - \kappa_n^2 <\mu^2$ where $\kappa_n$ satisfies 
\begin{eqnarray}
\mu = \mp \kappa_n \coth(\kappa_n L) \,\, {\tt (DL/DR)},
\end{eqnarray}
when the boundary condition (DL/DR) is imposed, respectively. When $|\mu L| \gg 1$, one can easily find an approximate analytic expression for the mass using the approximation $\coth(x)\simeq 1+2e^{-2x}$ for large $x\gg 1$: 
\begin{eqnarray}
m_1 \simeq 2 |\mu| e^{-|\mu L|} \,\,{\tt (DL/DR)},
\label{eq:small mass}
\end{eqnarray}
which explicitly shows the exponential suppression. On the other hand, all the other KK modes belong to the heavy branch and thus are hidden from the low energy regime when $R^{-1}$ is sufficiently large.

\begin{figure}[t]
\centering
\includegraphics[width=0.60\textwidth, angle=0]{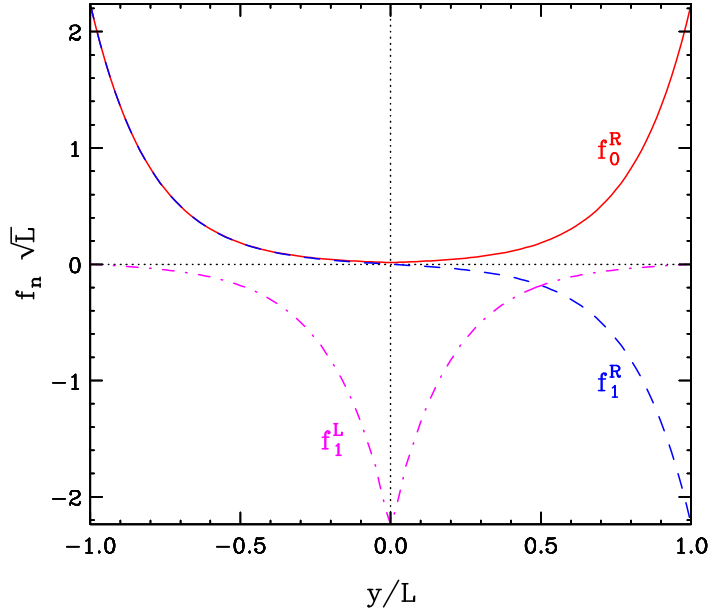}
\caption{The wave functions of the zero mode and the first Kaluza-Klein excitation modes with a large negative $\mu$. There is only the right-handed zero mode ($f^R_0$) by the orbifold boundary condition. A left-handed solution ($f^L_1$) forms a light ($m_1\simeq 2|\mu|e^{-|\mu L|}$) Dirac spinor with a right-handed partner ($f^R_1$). }
\label{fig:wave}
\end{figure}

Wave function profiles are shown in Fig.~\ref{fig:wave} for the massless zero mode and the two light level-1 modes which belong to the `light branch'.  These are {\it the only} wave profiles relevant for TeV scale physics in our setup since only the level-1 KK fermions can be light and accessible at current colliders when $R^{-1}\gg 1$ TeV. The first KK mode is $f^L_1$ (magenta, dot-dashed) which forms a Dirac spinor with $f^R_1$ (blue, dashed). The zero mode is $f^R_0$ (red, solid) which survives under the Dirichlet boundary condition for the left-handed state (DL). When $\mu L<-1$, the wave profile grows and is peaked at the boundaries. There is no $f_L^0$ mode because of the (DL) condition. Instead, $f^L_1$ behaves as an `would-be-zero-mode' solution as it approximately satisfies the Dirichlet boundary condition, $f^L_1 (y\to \pm L) \to 0$. This level-1 KK mode is exponentially light as is shown in Eq. \ref{eq:small mass}. 

\begin{figure}[t]
\centering
\includegraphics[width=0.60\textwidth, angle=0]{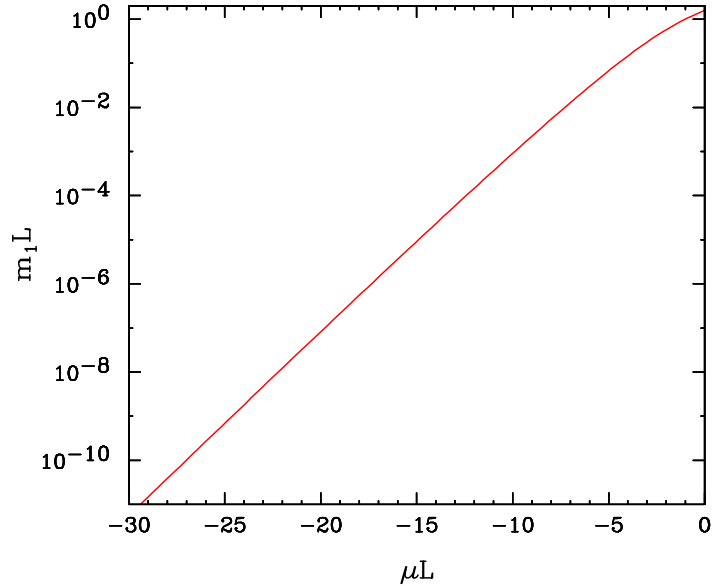}
\caption{$m_1 L$ as a function of $\mu L$. 
The $m_1 L$ decreases exponentially as the $\mu L$ increases.
Therefore the mass $m_1$ is sensitive to a small change in $\mu$ 
in the negative $\mu$ region of interest.}
\label{fig:mass}
\end{figure}

The masses of these light states are shown in Fig. \ref{fig:mass}, as a function of $\mu L$. 
For instance, if $L^{-1}=10^5$ GeV with the bulk mass $\sim 7.3 L^{-1}$, the mass of the light mode is about $1$ TeV. 
A more extreme case is GUT scale compactification, \eg,  $L^{-1}=10^{14}$ GeV. 
If the bulk mass parameter is about $29$ times larger than the compactification scale, 
the light mode is, again, about 1 TeV. 

In summary, if $\mu L $ has a large negative/positive value and (DL/DR) condition is imposed, a massless right-/left-handed zero mode and a light but massive level-1 KK state appear, respectively. All the higher KK fermionic states ($n_{\rm KK}\geq2$) and level-1 KK bosons are much heavier than these light states with a large mass gap ($\sim \max(|\mu|,R^{-1})$). 
Throughout this paper, we take the (DL) boundary condition by convention and hence the region of interest is the negative $\mu$ ($\mu L < -1$).

\section{Collider Implications}
\label{sec:collider}

We first extend the model in Ref. \cite{sued4} by introducing `right-handed' neutrinos \footnote{As the field resides in 5D, both the left- and right-handed component fields are included but we follow the conventional terminology.}, $N(x,y)$,  in addition to the Standard Model fermions in 5D; $N$ is a singlet under the SM interactions. By doing so, the model accommodates non-zero neutrino masses and provides a new potential DM candidate, the level-1 `right-handed' neutrino. 
Here we only show the new terms in the action which includes the Yukawa coupling with `right-handed' neutrinos: 
\begin{eqnarray}
S_n=\int d^4 x \int_{-L}^L dy \,\,\left[y\overline{\Psi}_\ell H ( N + e^{i\theta}N^c)+h.c.\right] \label{eq:action}. 
\end{eqnarray}
All the other interaction terms are given in Ref. \cite{sued4} and we follow the convention therein. 
The flavor indices are suppressed but the complex phase $\theta$ is explicitly shown. After the Higgs field develops the vev, the neutrino masses are generated. The complex phase, $\theta$, can lead to interesting phenomena such as CP-violation in the neutrino sector and deserves further study (see e.g., \cite{Maalampi}).

The collider phenomenology in this situation is found to be very different from MUED or 
ordinary SUED with a positive bulk mass.
First of all,  in principle, there are 6 different $\mu$'s for each generation; 
2 for the $SU(2)_W$-doublet quarks and leptons, $q=(q^u,q^d)$ and $\Psi_\ell=(\nu, \ell)$, 
4 for the four $SU(2)_W$-singlets, $u^c, d^c, e^c$ and $n$, where 
$u$ ($n$) and $d$ ($e$) represent up-type and down-type quarks (leptons), respectively.
Furthermore these $\mu$'s could be different for each of the three generations. 
Therefore the total number of the bulk masses is 18 
for three generations of quarks and leptons.
$R^{-1}$ controls the overall scale of all the other KK states.
As discussed in the previous section (see Fig.~\ref{fig:mass}), 
however, level-1 KK fermions can be found at the TeV scale even for a very large $R^{-1}$ 
by introducing negative bulk mass term for each of the corresponding fermions.
The flavor structure becomes richer as the variations of $\mu$ lead to the changes in wave function overlaps for the different flavors \cite{flavor}.
\begin{figure}[t]
\centering
\includegraphics[width=0.60\textwidth]{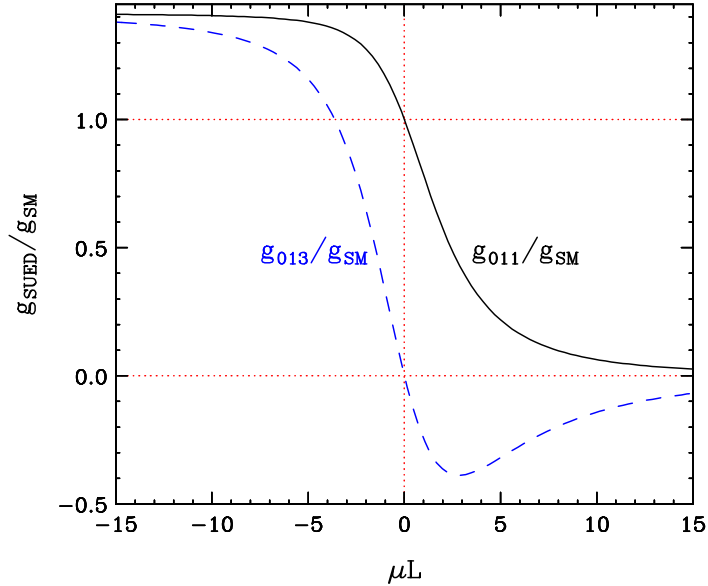}
\caption{The ratio of the tree-level coupling constant for a zero mode fermion ($f_0$)-level-1 KK fermion ($f^\prime_1$)-level-1 ($V_1$) (level-3 ($V_3$)) gauge boson and the corresponding SM coupling constant ($g_{011}/g_{SM}$ ($g_{013}/g_{SM}$))  is shown as a function of $\mu L \in (-15, 15)$. When $\mu$ vanishes, the ratio becomes 1(0) as expected for $g_{011}$ ($g_{013}$). }
\label{fig:g011}
\end{figure}

There are only two types of interactions which are relevant for the TeV scale 
because essentially only the $n_{KK}\leq 1$ states are relevant if we assume $R^{-1} \gg 1$ TeV.
The coupling for $f_1$-$\bar{f}_1^\prime$-$V_0$ is the same 
as the corresponding SM interaction, $f_0$-$\bar{f}_0^\prime$-$V_0$, 
while the coupling, $g_{011}$, for $f_1$-$\bar{f}_0^\prime$-$V_1$ (and its conjugate interaction) 
can be as large as $\sqrt{2}$ times the corresponding coupling strength ($g_{\rm SM}$) in the SM 
(see Fig.~\ref{fig:g011}, which completes the previous result for the positive $\mu$ in Ref. \cite{sued4}.). The behavior of $g_{011}/g_{\rm SM}$ in Fig.~\ref{fig:g011} is easily understood as follows. When $\mu$ becomes large and positive, the wave function profile of the zero mode fermion is essentially localized in the middle of the interval where the wave function of the first KK gauge boson vanishes. Thus the wave function overlap approaches the value zero. On the other hand, if the $\mu$ becomes large and negative, the wave function of the zero mode fermion tends to be localized near the boundaries. Then, the wave function normalization gives an additional factor of $\sqrt{2}$ 
beyond that of the $g_{SM}$ coupling.

The production mechanism of KK fermions is simpler than in MUED.
For example, {\it all} of the contributions discussed in Ref. \cite{Datta:2010us} 
that include KK levels higher than $n_{KK} \geq 2$ should be negligible.
Moreover all contributions arising from diagrams with KK bosons ($n_{KK}\geq 1$) 
can be ignored as well. Therefore the production of level-1 KK fermions is exactly the 
same as that of a vector-like fourth generation, {\ie} they are produced via 
SM gauge boson exchange only.

The decay of level-1 KK fermions, however, involves level-1 KK gauge bosons 
due to the conservation of KK-parity.
There are two types of decays;
$f_1 \to f_1^\prime V_0$ and 
$f_1 \to f_0^\prime V_1^\ast \to f_0^\prime \bar{f}_0^{\prime\prime} f_1^{\prime\prime\prime}$.
The former includes examples such as 
$\nu_1 \to n_1 h( n_1 Z)$, $\ell_1 \to \nu_1 W^-$, $\ell_1 \to n_1 W^-$, etc.
The latter case (three body decay via an offshell level-1 KK gauge boson) is more common. 
Loop-decays are even further suppressed partly due to KK-parity and 
partly because {\it all} KK bosons are heavy.

For further discussion let us assume the lightest KK particle (LKP) is the right-handed KK neutrino ($n_1$) and the next-to-LKP (NLKP) is the left-handed KK neutrino ($\nu_1$). The effect of the neutrino Yukawa coupling on the KK mass spectrum can be safely neglected.
As argued most fermions go through three body decays to the NLKP 
(the left-handed KK neutrino, $\nu_1$), and the partial decay width is given by 
\begin{equation}
\Gamma =\frac{G_{\tt eff}^2 M^5}{192 \pi^3} 
\Big (  1 - 8 \frac{m^2}{M^2}  + 8 \frac{m^6}{M^6} 
            - \frac{m^8}{M^8} - 24  \frac{m^4}{M^4} \log \big ( \frac{m}{M} \big )         \Big ) \, ,
\end{equation}
where $\frac{G_{\tt eff}}{\sqrt{2}} = \frac{g_{011}^2 R^2}{8}$, 
$m$ is the mass of the NLKP and $M$ is the mass of the decaying KK fermion.
The coupling changes as $\mu$ changes as shown in Fig.~\ref{fig:g011}.
In principle, there are contributions from interaction with higher KK gauge bosons. 
For example, a KK-number violating but KK-parity conserving interaction, 
$f_1$-$f^\prime_0$-$V_3$ ($f_1$-$f^\prime_0$-$V_{(2n_{KK}-1)}$ in general), 
can also mediate the above 3 body decay.
Including those higher KK modes, the $G_{\tt eff}$ becomes 
\begin{equation}
\frac{G_{\tt eff}}{\sqrt{2}} = \frac{g_{011}^2 R^2}{8} 
\Big \{ 1 + \frac{1}{9} \frac{g_{013}^2}{g_{011}^2} + 
\frac{1}{25} \frac{g_{015}^2}{g_{011}^2}+ \cdots    \Big \} \, .
\end{equation}
However the $V_3$ ($V_{(2n_{KK}-1)}$) is 3 ($2n_{KK}-1$) times 
heavier than $V_1$ resulting in a smaller decay amplitude by a factor of $9 = 3^2$ 
($(2n_{KK}-1)^2$).
Also the corresponding coupling $g_{013}$ ($g_{(2n_{KK}-1)}$) is smaller than $g_{011}$ 
in the region of our interest ($\mu L<-1$). 
Therefore the contribution from higher KK modes is negligible and we only include the first 
two terms in the $G_{\tt eff}$.

Fig.~\ref{fig:mu_vs_Rinv} shows the lifetime of the singlet KK lepton ($e_1$) 
as an illustration (see steeper lines) assuming it is the next-to-NLKP. 
In this case, the singlet KK lepton decays to 
the NLKP ($\nu_1$) via an offshell KK photon, 
$e_1 \to e \gamma_1^\ast \to e \bar{\nu} \nu_1 (e \nu \bar{\nu}_1)$ and the corresponding coupling 
replacement is $g_{SM}^2 \to \frac{g_Y^2 y_\ell y_e}{4}$ 
where the hypercharge for the doublet (singlet) is denoted by $y_{\ell}$ ($y_{e}$), respectively. 
For the purpose of demonstration, 
we fix the mass ratio $m/M$ to be 0.8 for our numerical results. 
The NLKP $\nu_1$ decays to the LKP $n_1$ 
(via $\nu_1 \to n_1 h$ or $\nu_1 \to n_1 Z$.). 
In collider experiments, most likely the NLKP is the missing particle 
since its lifetime is long enough to escape our detector 
(see discussion in the next section.). 
The $\tau = 10^{17}$ seconds line corresponds to the approximate age of our universe and 
the left side of this contour (`e') is not allowed. 
The smaller $\mu$ is in magnitude the shorter the lifetime.
In the region (`d') between $\tau \sim 1 {\rm ~sec~}$ and $ \tau < 10^{17}$ sec, 
one should be concerned about bounds arising 
from cosmology such as BBN due to the existence of long-lived charged particles.
The singlet KK lepton decays promptly (see region `a')
if the lifetime is shorter than $\sim 10^{-12}$ second which corresponds to the minimum length for the displaced vertex, $300 ~ {\rm \mu m} \sim \frac{10^{-12} }{\gamma\beta} {\rm ~sec}$, 
where the $\beta$ is the velocity of the produced particle and the $\gamma=1/\sqrt{1-\beta^2}$.
Therefore in region `b', the produced KK particle can leave tracks 
(either as a displaced vertex or as a long-lived CHArged Massive Particles (CHAMPs))
and decay inside the detector assuming the size of detector is about $20$ m.
In this case, one could directly measure the size of the extra dimension, 
$R$, by measuring the particle lifetime.
Finally in region `c', one can always observe CHAMPs since the $e_1$ escapes the detector and  decays far outside.

The other two lines labeled by `100 GeV' and `1 TeV' show constant contours of 
KK mass in the same $R^{-1}$-$\mu L$ plane.
Considering contours of masses and lifetime together, one can see that 
$R^{-1}$ can not be larger than $\sim 10^5$ GeV, for a prompt decay.
For TeV fermions, $R^{-1} \gtrsim 10^{12}$ GeV is ruled out since the decaying particle 
is cosmologically stable ($\tau \gtrsim 10^{17}$ sec.).
\FIGURE[t]{
\centerline{
\includegraphics[width=0.60\textwidth]{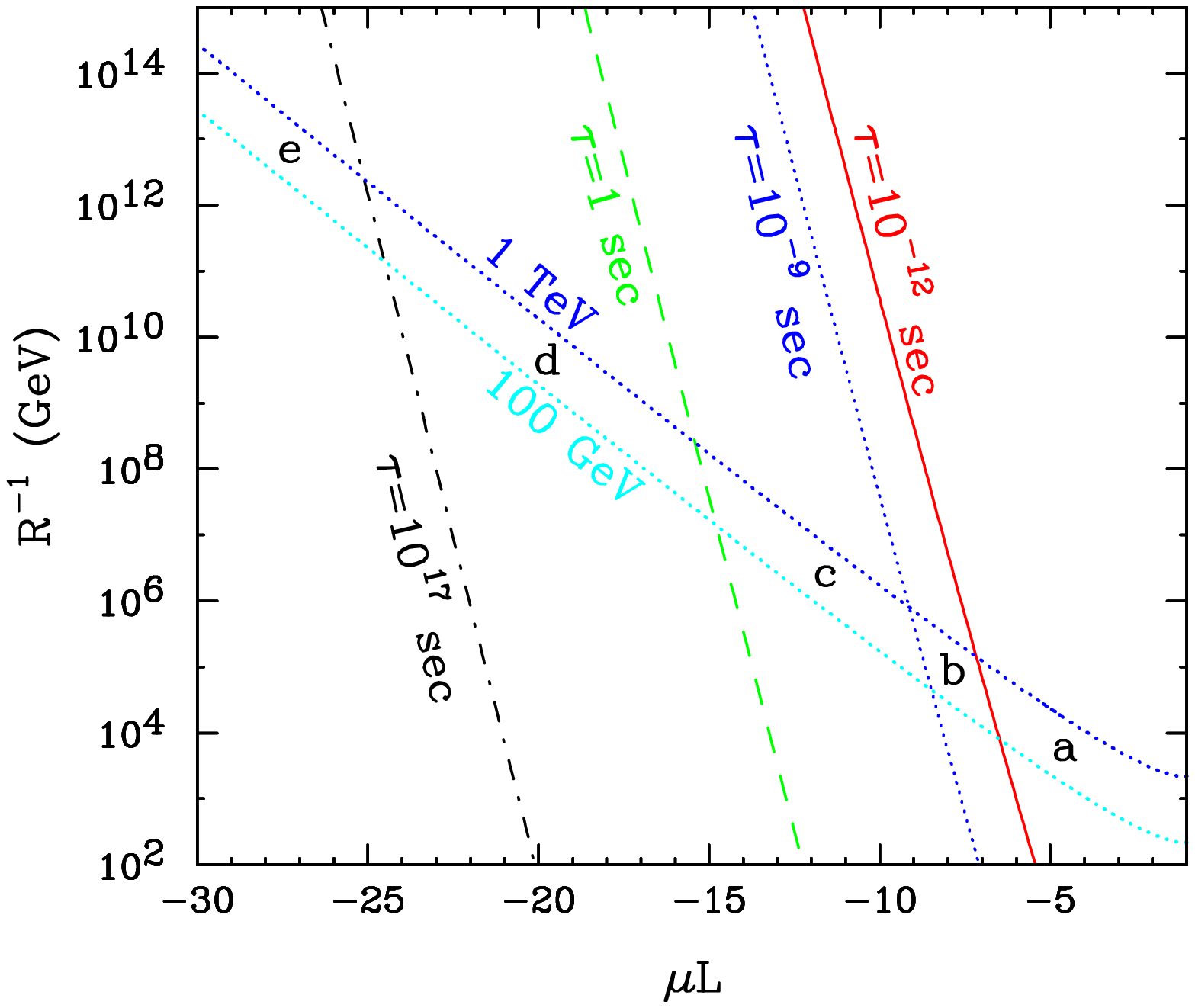} }
\caption{The lifetime for the decay of the singlet KK lepton into $\nu_1$ via the offshell KK photon in the $\mu L$-$R^{-1}$ plane. The mass ratio $m/M$ is taken to be 0.8. 
The lifetime of $\nu_1$ is $\tau_{\nu_1} \lesssim 1$ second in the 
parameter space region consistent with WMAP (see Section \ref{sec:dm}.).}
\label{fig:mu_vs_Rinv}}

The third generation KK quarks lead to more interesting signatures.
The doublet KK top decays via $q^t_1 \to q^b W_1^{+\ast} \to q^b \bar{\ell} \nu_1$ and 
the doublet KK bottom goes through $q^b_1 \to q^t W_1^{-\ast} \to q^t \ell \bar{\nu}_1$ 
($q^t_1 \to q^b_1 W^+$ and $q^b_1 \to q^t_1 W^-$ are also possible, if kinematically allowed.), 
while the pair production of the singlet KK top ($t_1$) would give rise to $t\bar{t} + \met$, 
where $t_1 \to t \gamma^\ast_1 \to t \nu \bar{\nu}_1 (t \bar{\nu} \nu_1)$. 
The recent developments \cite{Konar:2009qr,Konar:2009wn,Burns:2008va} employing 
the kinematic variable $M_{T2}$ 
can be very useful for determination of KK fermion masses in such topologies.

\section{Dark matter}
\label{sec:dm}

In general there are several dark matter candidates in UED. 
Examples include the KK photon \cite{Servant:2002aq,Cheng:2002ej}, KK neutrino \cite{Servant:2002aq}, KK Z \cite{Arrenberg:2008wy}, KK Higgs \cite{Cembranos:2006gt}, KK graviton \cite{Shah:2006gs}, spinless photon \cite{Dobrescu:2007ec} etc. In our setup, on the other hand,  the mass scale of KK bosons (including the KK photon, KK Z, KK Higgs, KK graviton, spinless photon etc) is large ($\sim R^{-1}\gg 1\,{\rm TeV}$) and hence the only viable candidates are KK neutrinos: one in $\Psi_\ell=(\nu, \ell)_L$ and the other one in $N$. The left-handed neutrino sitting in the doublet $\Psi_\ell$, $\nu_1$,  is known to have a problem with direct detection constraints due to the large elastic scattering cross section arising from $Z$ exchange \cite{Servant:2002aq}.
The first KK state, $n_1$, of $N$, on the other hand, can be a good dark matter candidate as has been shown in Ref. \cite{Matsumoto:2007dp}. Here we will show that it can be a good dark matter candidate for a different reason.

\begin{figure}[t]
\centering
\includegraphics[width=0.60\textwidth]{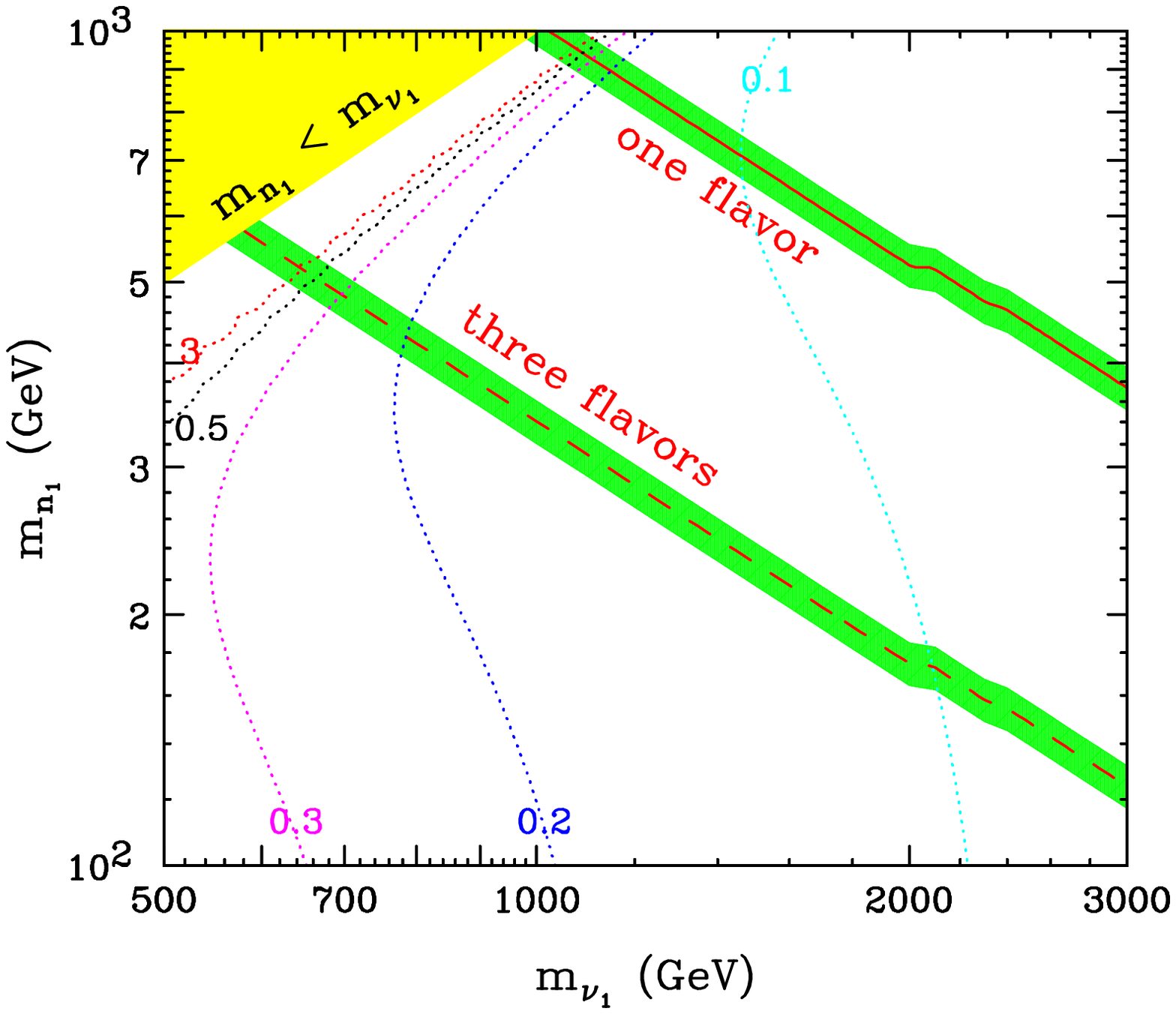}
\caption{The abundance of the right-handed KK neutrino ($n_1$) is shown in red curves 
(solid for `one flavor' of DM and dashed for `three flavors' of DM) 
and the green band represents 7-year WMAP allowed region accounting for all of dark matter. 
In principle, the area below the green band is also allowed. 
The dotted curves show constant contours for the lifetime of $\nu_1$. 
The yellow-shaded region in the left-upper corner is kinematically forbidden.}
\label{fig:mLKP_vs_mNLKP}
\end{figure}

For the right-handed KK neutrino ($n_1$) being dark matter candidate, 
there are three contributions to its relic abundance \cite{Matsumoto:2007dp}.
In the early universe, when level-1 ($n_{KK}=1$) and higher right-handed KK neutrinos ($n_{KK} \geq 2$) are 
produced in the thermal equilibrium, they remain intact due to the very weak Yukawa interaction.
The other KK particles remain in the thermal bath until the temperature reaches $T \sim 1/(20 R)$.
Then in the late universe ($T \ll 1/R$), the first right-handed KK neutrinos (LKPs) are produced 
from the decay of the higher KK neutrinos and the next-to-lightest KK particle (NLKP). 
As a result, the relic abundance of the LKP is determined by three processes \cite{Matsumoto:2007dp} 
\begin{equation}
\Omega = \Omega_{\rm thermal} + \Omega_{{n_{KK} \geq 2}~ \rm decay} + \Omega_{\rm NLKP ~decay} \, .
\end{equation}
However it is known that for $m_\nu \gsim 0.3 $ eV, 
the first two contributions are negligible.
(Also $m_{n_{KK}\geq 2} \gsim \frac{2}{R} \gg m_{n_1}$.)
Therefore we expect that the dominant contribution to $\Omega$ to be from the late decay of the NLKP after freeze-out. The abundance of the right-handed KK neutrino ($n_1$) is shown in Fig.~\ref{fig:mLKP_vs_mNLKP} 
as red curves (solid for `one flavor' of DM and dashed for `three flavors' of DM). 

Here we take $\nu_1$ as the NLKP 
\footnote{For the charged lepton ($\ell_1$) as the NLKP, similar results are obtained.
The decay $\ell_1 \to {n}_1  W^-$ is similar 
to the process $\nu_1 \to n_1 Z$ with appropriate replacement in the couplings.}.
We then repeat a similar analysis as in Ref. \cite{Servant:2002aq} but now 
assume that all other KK particles at level-1 except for the left-handed charged KK leptons, $\ell_1$, are heavy so that they do not participate 
in the annihilation processes (see Refs. \cite{Kong:2005hn,Burnell:2005hm} 
for the effects of coannihilation.). 
It is important to include coannihilation with $\ell_1$ 
since the mass splitting with $\nu_1$ 
is expected to be small due to the same bulk mass appearing 
for both members of the $SU(2)_W$-doublet, 
$(\nu_1$, $\ell_1)$. 
Once NLKPs decouple from the thermal bath, they decay into the LKP shortly 
after the freeze-out, and 
the abundance of the LKP is given by 
\begin{equation}
\Omega_{LKP} = \frac{m_{LKP}}{m_{NLKP}} \Omega_{NLKP} \, .
\end{equation}
This is very similar to the SuperWIMP scenario (see \cite{Feng:2010tg} for non-WIMP dark matter) 
except for the scale of the NLKP lifetime.
The decay width of a typical SuperWIMP is suppressed by the Planck mass while in our case, it is suppressed by the small Yukawa coupling in the neutrino sector. This difference explains the different lifetime of the NLKP, which is shown as the dotted contours in Fig.~\ref{fig:mLKP_vs_mNLKP}.
The dominant decay modes are $\nu_1 \to n_1 h$ and $\nu_1 \to n_1 Z$.
The partial width of $\nu_1 \to n_1 h$ is given by 
\begin{eqnarray}
\Gamma^{\nu_1}_{ n_1 h} &=& 
\frac{y^2 \lambda \Big ( \frac{\mr}{\ml}, \frac{\mh}{\ml} \Big )}{16 \pi \ml} 
                                    \Big ( (\ml+\mr)^2 - \mh^2 \Big ) , ~~~~
\end{eqnarray}
where $\lambda^2(x,y) = 1+ x^4 + y^4 -2 x - 2y - 2 x y$ and the $y$ is the Yukawa coupling 
in Eq. \ref{eq:action}.
The partial width of $\nu_1 \to n_1 Z$ is
\begin{eqnarray}
\Gamma^{\nu_1}_{n_1 Z} &=& 
\frac{ g_Z^2\sin^2\theta\lambda \Big ( \frac{\mr}{\ml}, \frac{\mz}{\ml} \Big )}{16\pi \ml}
\Big [ \mll + \mrr - m_Z^2 
- 6 \ml  \mr  \\
&& \hspace*{4cm}+ \frac{( \mll- \mrr - m_Z^2  )(\mll - \mrr + m_Z^2)}{m_Z^2}
\Big ]
 , \nonumber
\end{eqnarray}
where $g_Z = \frac{e}{2 s_W c_W}$, $\sin\theta = \frac{m_\nu}{\ml-\mr}$ and we take 
$m_\nu = 0.053$ eV and $m_h=120$ GeV for numerical purposes.
In most of parameter space, the NLKP decays within $\sim 1$ 
second which is short enough for successful 
big bang nucleosynthesis to be maintained. 

\section{Conclusion}
\label{sec:conclusion}

In this paper we have discussed the possibility 
that the level-1 KK fermions are much lighter than the corresponding KK bosons.
All other KK states  ($n_{KK}\geq 2$) can be completely decoupled 
from the low energy ($\sim$TeV) physics.
This can be naturally realized in Split-UED with large negative bulk fermion masses.
Such a scenario would look very similar 
to a model with extra generations with a discrete symmetry.
A natural dark matter candidate is the `right-handed' KK neutrino, 
whose relic density is mostly given by the density of NLKP.
This is similar to a SuperWIMP scenario but the associated mass 
scale is determined by the SM neutrino masses, which is small compared to TeV scale. 
However this suppression is not as large as is the $M_{Pl}^{-1}$ coupling 
in the SuperWIMP case, and 
hence the lifetime of NLKP tends to be short enough 
(typically shorter than $\sim{\cal O}(1)$ second) to avoid astrophysical or cosmological constraints such as BBN.

There could be up to 6 (= 3 generations $\times$ 2 chiral states) extra vector-like generations (each pair corresponding to the level-1 KK state of the SM generation, since there are two KK fermions for each SM fermion), depending on the scale of the corresponding bulk masses.
Most of them will decay through three body process to the NLKP and their lifetime depends on the values of  $R^{-1}$ and the bulk masses.  A model of a fourth generation with a discrete symmetry can lead to interesting dark matter and collider phenomenology.

Finally, we comment on constructing the minimal SUED model assuming a universal bulk mass for all fermions. In this case, the mass splitting between the states can be obtained from RG running between TeV and the cutoff scale, which is a high scale larger than $R^{-1}$, where the masses are universal. MUED is recovered when the universal mass term vanishes. If there is a big gap between the first KK scale and the higher ones, as is considered in this paper, the effective theory below $R^{-1}$ includes extra vector-like generations and nothing else. The long RG running in this case is only logarithmic below $R^{-1}$.

\bigskip

\acknowledgments
S. C. Park is supported by the World Premier International Research Center Initiative 
(WPI initiative) by MEXT and also supported by the Grant-in-Aid for scientific 
research (Young Scientists (B) 21740172) from JSPS, Japan.
K. Kong and T. G. Rizzo are supported in part by the DOE under contract DE-AC02-76SF00515.


\begin{thebibliography}{999}


\bibitem{ADD1}
  N.~Arkani-Hamed, S.~Dimopoulos and G.~R.~Dvali,
  ``The hierarchy problem and new dimensions at a millimeter,''
  Phys.\ Lett.\  B {\bf 429}, 263 (1998)
  [arXiv:hep-ph/9803315].

\bibitem{RS}
  L.~Randall and R.~Sundrum,
  ``A large mass hierarchy from a small extra dimension,''
  Phys.\ Rev.\ Lett.\  {\bf 83}, 3370 (1999)
  [arXiv:hep-ph/9905221].

\bibitem{Appelquist:2000nn}
  T.~Appelquist, H.~C.~Cheng and B.~A.~Dobrescu,
  ``Bounds on universal extra dimensions,''
  Phys.\ Rev.\  D {\bf 64}, 035002 (2001)
  [arXiv:hep-ph/0012100].



\bibitem{Antoniadis}
  I.~Antoniadis,
  ``A Possible new dimension at a few TeV,''
  Phys.\ Lett.\  B {\bf 246}, 377 (1990).

\bibitem{Cheng:2002ab}
  H.~C.~Cheng, K.~T.~Matchev and M.~Schmaltz,
  ``Bosonic supersymmetry? Getting fooled at the CERN LHC,''
  Phys.\ Rev.\  D {\bf 66}, 056006 (2002)
  [arXiv:hep-ph/0205314].

\bibitem{Cheng:2002iz}
  H.~C.~Cheng, K.~T.~Matchev and M.~Schmaltz,
  ``Radiative corrections to Kaluza-Klein masses,''
  Phys.\ Rev.\  D {\bf 66}, 036005 (2002)
  [arXiv:hep-ph/0204342].


\bibitem{sued1}
  S.~C.~Park and J.~Shu,
  ``Split-UED and Dark Matter,''
  Phys.\ Rev.\  D {\bf 79}, 091702(R) (2009)
  [arXiv:0901.0720 [hep-ph]].
  
\bibitem{sued2}
  C.~R.~Chen, M.~M.~Nojiri, S.~C.~Park, J.~Shu and M.~Takeuchi,
  ``Dark matter and collider phenomenology of split-UED,''
  JHEP {\bf 0909}, 078 (2009)
  [arXiv:0903.1971 [hep-ph]].
  
\bibitem{sued3}
  C.~R.~Chen, M.~M.~Nojiri, S.~C.~Park and J.~Shu,
  ``Kaluza-Klein Dark Matter After Fermi,''
  arXiv:0908.4317 [hep-ph].
    
\bibitem{sued4}
  K.~Kong, S.~C.~Park and T.~G.~Rizzo,
  ``Collider Phenomenology with Split-UED,''
    JHEP {\bf 04} (2010) 081
  [arXiv:1002.0602 [hep-ph]]. 



\bibitem{Servant:2002aq}
G.~Servant and T.~M.~Tait,
``Is the lightest Kaluza-Klein particle a viable dark matter candidate?,''
Nucl.\ Phys.\ B {\bf 650}, 391 (2003)
[arXiv:hep-ph/0206071].

\bibitem{Cheng:2002ej}
  H.~C.~Cheng, J.~L.~Feng and K.~T.~Matchev,
  ``Kaluza-Klein dark matter,''
  Phys.\ Rev.\ Lett.\  {\bf 89}, 211301 (2002)
  [arXiv:hep-ph/0207125].


\bibitem{ArkaniHamed:1999dc}
  N.~Arkani-Hamed and M.~Schmaltz,
  ``Hierarchies without symmetries from extra dimensions,''
  Phys.\ Rev.\  D {\bf 61}, 033005 (2000)
  [arXiv:hep-ph/9903417].



\bibitem{Datta:2010us}
  A.~Datta, K.~Kong and K.~T.~Matchev,
  ``Minimal Universal Extra Dimensions in CalcHEP/CompHEP,''
  arXiv:1002.4624 [hep-ph].



\bibitem{Konar:2009qr}
  P.~Konar, K.~Kong, K.~T.~Matchev and M.~Park,
  ``Dark Matter Particle Spectroscopy at the LHC: Generalizing MT2 to
  Asymmetric Event Topologies,''
  arXiv:0911.4126 [hep-ph].

\bibitem{Konar:2009wn}
  P.~Konar, K.~Kong, K.~T.~Matchev and M.~Park,
  ``Superpartner mass measurements with 1D decomposed MT2,''
  arXiv:0910.3679 [hep-ph].

\bibitem{Burns:2008va}
  M.~Burns, K.~Kong, K.~T.~Matchev and M.~Park,
  ``Using Subsystem MT2 for Complete Mass Determinations in Decay Chains with
  Missing Energy at Hadron Colliders,''
  JHEP {\bf 0903}, 143 (2009)
  [arXiv:0810.5576 [hep-ph]].



\bibitem{Arrenberg:2008wy}
  S.~Arrenberg, L.~Baudis, K.~Kong, K.~T.~Matchev and J.~Yoo,
  ``Kaluza-Klein Dark Matter: Direct Detection vis-a-vis LHC,''
  Phys.\ Rev.\  D {\bf 78}, 056002 (2008)
  [arXiv:0805.4210 [hep-ph]].

\bibitem{Cembranos:2006gt}
  J.~A.~R.~Cembranos, J.~L.~Feng and L.~E.~Strigari,
  ``Exotic collider signals from the complete phase diagram of minimal
  universal extra dimensions,''
  Phys.\ Rev.\  D {\bf 75}, 036004 (2007)
  [arXiv:hep-ph/0612157].

\bibitem{Shah:2006gs}
  N.~R.~Shah and C.~E.~M.~Wagner,
  ``Gravitons and dark matter in universal extra dimensions,''
  Phys.\ Rev.\  D {\bf 74}, 104008 (2006)
  [arXiv:hep-ph/0608140].

\bibitem{Dobrescu:2007ec}
  B.~A.~Dobrescu, D.~Hooper, K.~Kong and R.~Mahbubani,
  ``Spinless photon dark matter from two universal extra dimensions,''
  JCAP {\bf 0710}, 012 (2007)
  [arXiv:0706.3409 [hep-ph]].


\bibitem{Matsumoto:2007dp}
  S.~Matsumoto, J.~Sato, M.~Senami and M.~Yamanaka,
  ``Relic abundance of dark matter in universal extra dimension models with
  right-handed neutrinos,''
  Phys.\ Rev.\  D {\bf 76}, 043528 (2007)
  [arXiv:0705.0934 [hep-ph]].

\bibitem{Kong:2005hn}
  K.~Kong and K.~T.~Matchev,
  ``Precise calculation of the relic density of Kaluza-Klein dark matter in
  universal extra dimensions,''
  JHEP {\bf 0601}, 038 (2006)
  [arXiv:hep-ph/0509119].

\bibitem{Burnell:2005hm}
  F.~Burnell and G.~D.~Kribs,
  ``The abundance of Kaluza-Klein dark matter with coannihilation,''
  Phys.\ Rev.\  D {\bf 73}, 015001 (2006)
  [arXiv:hep-ph/0509118].

\bibitem{Feng:2010tg}
  J.~L.~Feng,
  ``Non-WIMP Candidates,''
  arXiv:1002.3828 [hep-ph].
  
\bibitem{Maalampi}
  J.~Maalampi, I.~Vilja and H.~Virtanen,
  ``Thermal leptogenesis in a 5D split fermion scenario with bulk neutrinos,''
  arXiv:0912.4377 [hep-ph].

\bibitem{flavor}
 C.~Cs\'aki, J. ~ Heinonen, J.~ Hubisz,
S.~C.~Park and J.~Shu, in preparation.



\end{thebibliography}
\end{document}